\documentclass[12pt]{iopart}
\usepackage{epsfig}
\usepackage{graphicx}
\pdfoutput=1

\usepackage{iopams}  
\begin{document}

\title[Lagrangian chaos in an ABC--Dynamo]{Lagrangian chaos in an ABC--forced nonlinear dynamo}

\author{Erico L. Rempel$^1$,Abraham C.--L. Chian$^{2,3}$ and Axel Brandenburg$^{4,5}$}

\address{$^1$ Institute of Aeronautical Technology (ITA), World Institute
for Space Environment Research (WISER), S\~ao Jos\'e dos Campos -- SP 12228--900, Brazil\\

$^2$ Observatoire de Paris, LESIA, CNRS, 92190 Meudon, France\\

$^3$ National Institute for Space Research (INPE) and World Institute for Space Environment Research (WISER), 
P.O. Box 515, 12227-010 S\~ao Jos\'e dos Campos-SP, Brazil\\

$^4$ NORDITA, AlbaNova University Ctr, Stockholm, Sweden\\ 

$^5$ Department of Astronomy, Stockholm University, SE-10691 Stockholm, Sweden
}
\ead{rempel@ita.br}

\begin{abstract}

The Lagrangian properties of the velocity field in a magnetized fluid 
are studied using three-dimensional simulations of a helical magnetohydrodynamic dynamo. 
We compute the attracting and repelling Lagrangian coherent structures, which are dynamic lines and surfaces in the velocity field 
that delineate particle transport in flows with chaotic streamlines and act as transport barriers.  
Two dynamo regimes are explored, one with a robust coherent mean magnetic field and one with intermittent 
bursts of magnetic energy. The Lagrangian coherent structures and the statistics of finite--time Lyapunov exponents indicate
that the stirring/mixing properties of the velocity field decay as a linear function of the magnetic energy.
The relevance of this study for the solar dynamo problem is discussed.

\end{abstract}

\maketitle

\section{Introduction}

Transport in chaotic flows is governed by a combination of
stirring and diffusion. Stirring refers to the transport, stretching, twisting and folding of fluid elements
and, consequently, of scalar or vector quantities advected by the flow, such
as temperature, light particles or magnetic field lines in a magnetohydrodynamic (MHD) system.
This process creates complex tracer patterns in the flow, including filaments and sheets, as the fluid elements are
deformed in different directions. Diffusion is responsible for homogenizing the distribution
of tracers and blurring the patterns created by the chaotic stirring, being usually more important in small scales \cite{lekien07}.  
This paper deals with the problem of chaotic stirring in magnetized flows. Throughout the paper we 
use the terms ``stirring" and ``mixing" interchangeably. 

Passive scalars are quantities that are passively advected by the flow, i.e., their back--reaction on the advecting velocity field is disregarded. 
They constitute
a powerful way to study transport in hydrodynamical and MHD
flows (for a review, see Falkovich et al. \cite{falkovich01}). 
We employ passive scalars to investigate how the magnetic field can affect particle transport 
and the stirring/mixing properties of a velocity field in MHD simulations through the Lorentz force. 
We adopt direct numerical simulations of resistive three--dimensional (3--D) compressible MHD equations
with a helical forcing, which has been used elsewhere as a prototype of the $\alpha^2$ dynamo model 
of mean field dynamo theory \cite{rempel09,rempel11}. 
 
In the Lagrangian approach to turbulent transport the dynamics of fluids is studied by following the trajectories
of a large number of fluid elements or tracer particles. The specific trajectories of individual particles
are not very useful in this type of investigation in chaotic flows, since sensitivity to initial conditions
means that particles that are arbitrarily close may experience exponential divergence with time.
However, it is possible to detect certain material lines in the flow that repel or attract
fluid elements. These
repelling and attracting material lines are time--dependent analogous to the stable
and unstable manifolds of hyperbolic fixed points in dynamical systems theory and form 
transport barriers in flows with chaotic streamlines, being called Lagrangian coherent structures (LCS).
The LCS have been used to describe hydrodynamic turbulence in 3--D numerical simulations \cite{green07},
laboratory experiments \cite{voth02,mathur07} and observational data of oceans \cite{sandulescu07,olascoaga08} 
and the atmosphere \cite{tang10},
as well as 2--D numerical simulations of magnetized fusion plasmas \cite{padberg07}, magnetic reconnection \cite{borgogno2011},
and 3--D MHD simulations of conservative \cite{leoncini2006} and dissipative \cite{rempel11} fields.

One of the most widely used Lagrangian tools are the finite-time Lyapunov exponents (FTLE), also known as direct Lyapunov
exponents. The FTLE are a measure of local chaos and quantify the dispersion of particles in a region of the flow
during a finite time.
In the context of dynamo theory, the stretching rate of material lines in a fluid
can be used to explain the amplification of magnetic fields by the mechanism of stretch--twist--fold 
dynamo \cite{childress95}. 
Examples of applications of the FTLE in dynamo simulations include the growth of seed magnetic fields in the 
kinematic dynamo problem \cite{galloway_frisch86,galloway_proctor92,smith04}, nonlinear MHD dynamos 
\cite{axel95,cattaneo96,zienicke98} and the amplification
of interstellar magnetic fields and turbulent mixing by supernova--driven turbulence in compressible MHD simulations \cite{balsara05}.
It was shown by Haller \cite{haller01} that FTLE can also be used to 
identify repelling and attracting Lagrangian coherent structures.

We present the detection of Lagrangian coherent structures
for two different dynamo regimes in the 3--D compressible MHD equations with the isotropic and helical ABC forcing. We focus
on the change in transport and mixing properties of the flow when the system undergoes a transition 
whereby a large--scale spatially coherent magnetic field loses its stability. The transition, which
occurs after an increase in the magnetic diffusivity, results in strongly intermittent time series of magnetic energy. 
In the intermittent regime, the lower magnetic energy causes chaotic mixing 
to increase, resulting in higher
stretching rates of material lines. Chaotic mixing is quantified by 
the finite--time Lyapunov exponents, which show a linear dependence on the magnetic energy. In section II of this paper
we define Lagrangian coherent structures and how they relate to chaotic stirring in fluids;
Section III describes the model of MHD dynamo adopted; the numerical analysis is discussed in section IV and section
V presents the conclusions and possible ways to apply our techniques to observational data of the solar dynamo.

\section{Lagrangian Coherent Structures}

Let $D\subset\mathbb{R}^{3}$ be the domain of the fluid to be studied,
let $\mathbf{x}(t_{0})\in D$ denote the position of a passive particle
at time $t_{0}$ and let $\mathbf{u}(\mathbf{x},t)$ be the velocity
field defined on $D$. The motion of the particle is given by the
solution of the initial value problem 

\begin{equation}
\mathbf{\dot{x}}=\mathbf{u}(\mathbf{x},t), \qquad \mathbf{x}(t_{0})=\mathbf{x}_{0}.
\label{eq odes}
\end{equation}

Let us define the following flow map: $\phi_{t_{0}}^{t_{0}+\tau}:\mathbf{x}(t_{0})\mapsto\mathbf{x}(t_{0}+\tau)$.
The deformation gradient is given by $J=\mathrm{d}\phi_{t_{0}}^{t_{0}+\tau}(\mathbf{x})/\mathrm{d}\mathbf{x}$
and the finite-time right Cauchy-Green deformation tensor is given
by $\triangle=J^{T}J$. Let $\lambda_{1}>\lambda_{2}>\lambda_{3}$
be the eigenvalues of $\triangle$. Then, the finite-time Lyapunov
exponents or direct Lyapunov exponents of the trajectory of
the particle are defined as \cite{shadden05}:

\begin{equation}
\sigma_{i}^{t_{0}+\tau}(\mathbf{x})=\frac{1}{|\tau|}\ln\sqrt{\lambda_{i}},\qquad i=1,2,3.
\end{equation}


The maximum FTLE gives the finite-time average of the maximum rate
of divergence or stretching between the trajectories of a fiducial
particle at $\mathbf{x}$ and its neighboring particles. The maximum
stretching is found when the neighboring particle $\mathbf{y}$ is
such that $\delta\mathbf{x}=\mathbf{x}-\mathbf{y}$ is initially aligned
with the eigenvector of $\triangle$ associated with $\lambda_{1}$.
A positive $\sigma_{1}$ is the signature of chaotic streamlines in
the velocity field. The other exponents provide information about
stretching/contraction in other directions and can be useful to interpret
the local dynamics of the fluid. In an ideal conductive fluid, the frozen-in
condition implies that a magnetic line aligned with an infinitesimal vector connecting 
two close fluid elements will evolve as this vector \cite{zeldovich84}.
As pointed out by Balsara et al. \cite{balsara05}, for finite resistivity
and compressible flows, flow regions with three positive Lyapunov exponents expand in all
three directions and tend to dilute out the magnetic field; regions
with two positive Lyapunov exponents and one negative exponent tend
to concentrate the magnetic fields into sheet--like structures; regions
with one positive and two negative exponents tend to mold the magnetic
fields into filamentary structures; compression in all directions
is found when all exponents are negative. 
On the other hand, local minima in the maximum FTLE field might provide a way to
detect the position of the center of vortices in the velocity field, since vortices may
be viewed as material tubes of low particle dispersion \cite{cucitore99}. 

\begin{figure*}[ht]
\begin{center}
 \includegraphics[width=0.5\columnwidth]{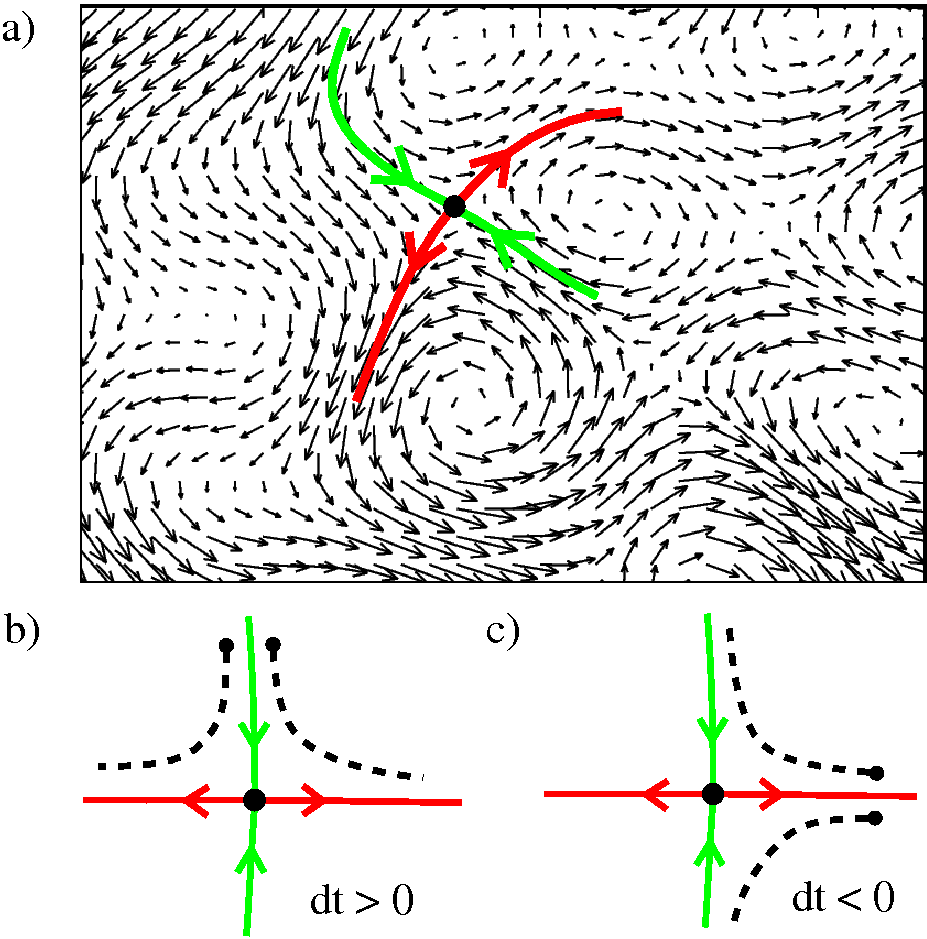}
\end{center}
 \caption{\label{fig saddless} (Color online) (a) Velocity vector field with a saddle point (black circle) and its stable (green) and unstable (red)
manifolds; (b) schematic drawing of the forward--time trajectories (dashed lines) of two passive particles that straddle the stable manifold of a saddle point;
(c) schematic drawing of the backward--time trajectories (dashed lines) of two passive particles that straddle the unstable manifold of a saddle point. }
 \end{figure*}

Finite-time Lyapunov exponents are also useful to detect attracting and repelling material
lines that act as barriers to particle transport in the velocity field.
A material line is a smooth curve of fluid particles advected by the velocity field \cite{haller01}.
These attracting and repelling material lines are the analogous of stable and unstable manifolds
of time-independent fields. The study of 2--D flows is helpful to understand
the role of material lines. Consider a 2--D steady flow, where the velocity field
does not change with time. In the presence of counter--rotating vortices, hyperbolic (saddle) points are expected to
be found, such as the one illustrated in Fig. \ref{fig saddless}(a). The trajectories of passive scalars
follow the velocity vectors in the vicinity of the hyperbolic point. Thus, particles lying on the stable manifold (green line)
are attracted to the saddle point in the forward--time dynamics and trajectories on the unstable manifold (red
line) converge to the saddle point in the backward--time dynamics. 

Two particles are said to straddle a manifold if the line segment 
connecting them crosses the manifold. The maximum FTLE has particularly high values on the stable manifold
in forward--time, since nearby trajectories straddling the manifold will experience exponential divergence when they approach 
the saddle point, as shown in Fig. \ref{fig saddless}(b). Similarly, the FTLE field exhibits a local maximizing curve ({\it ridge})
along the unstable manifold in backward--time dynamics, since trajectories straddling the unstable manifold
diverge exponentially when they approach the saddle point in reversed--time, as in Fig. \ref{fig saddless}(c).
Thus, ridges in the forward--time FTLE field mark the stable manifolds of hyperbolic points and ridges in the backward--time FTLE field
mark the unstable manifolds.

Analogously, for a time-dependent velocity field, regions of maximum material
stretching generate ridges in the FTLE field. 
Thus, repelling material lines (finite-time stable manifolds) produce 
ridges in the maximum FTLE field in the forward-time system  
and attracting material lines 
(finite-time unstable manifolds) produce ridges in the backward-time system
\cite{padberg07,haller01,shadden05}. These material 
lines are called Lagrangian coherent structures (LCS).

\begin{figure*}[ht]
\begin{center}
 \includegraphics[width=0.5\columnwidth]{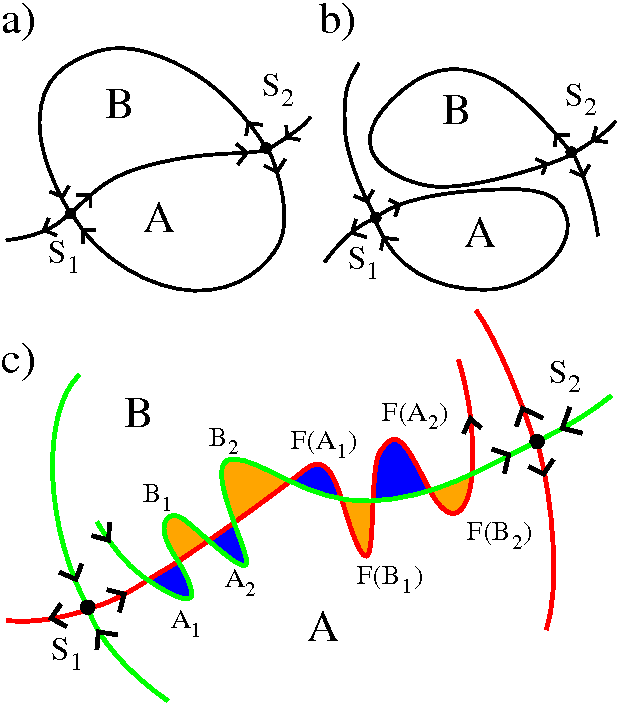}
\end{center}
 \caption{\label{fig tangles} (Color online) (a) Heteroclinic connection between saddle points
$S_1$ and $S_2$. There can be no transport of scalars between regions $A$ and $B$. (b) Homoclinic connections.
Again, there is no transport between regions $A$ and $B$.
(c) Heteroclinic tangle between the unstable manifold of $S_1$ and the stable manifold of $S_2$. The transport of particles between
regions $A$ and $B$ is possible through lobe dynamics.}
 \end{figure*}

Stable and unstable manifolds are invariant sets, which means that a particle on the manifold
will stay on it for all time. They form natural barriers to transport between different regions of a fluid,
as seen in Figure \ref{fig tangles}.
Figure \ref{fig tangles}(a) shows a topological configuration where one branch of the unstable manifold of 
the saddle point $S_1$ smoothly joins the stable manifold of another saddle point $S_2$ in a heteroclinic connection.
Simultaneously, the two branches of the unstable manifold of $S_2$ are connected to the stable manifold of $S_1$,
enclosing regions $A$ and $B$. Particles trapped in $A$ or $B$ cannot cross the barriers formed by the manifolds, 
since these are invariant sets. Figure \ref{fig tangles}(b) shows another type of trapping region, formed by a homoclinic connection,
where one branch of the unstable manifold of a saddle point joins its own stable manifold. Trajectories in regions $A$ and $B$ usually
circulate around a focus, as the manifolds mark the borders of vortices in the velocity field.
Transport between different vortices is only possible when there is a transversal crossing between stable and 
unstable manifolds, through a mechanism called lobe dynamics \cite{rom-kedar90,wiggins05}.
It is easier to understand this mechanism with a periodic flow. Suppose that the velocity field is time--dependent
but periodic, such that $\mbox{\bf u}(\mbox{\bf x},t) = \mbox{\bf u}(\mbox{\bf x},t+T)$, where $T$ is the period.
Let $F$ be the stroboscopic Poincar\'e map defined by $F(\mbox{\bf x}(t))=\phi_{t}^{t+T}(\mbox{\bf x})$. There are still points where
the velocity is instantly zero, but now they are moving. Since these points are not fixed, they are called stagnation
points. After $T$ time units a stagnation point will return to its original position. Therefore, under the map $F$ a stagnation point 
is seen as a fixed point. Figure \ref{fig tangles}(c) shows a heteroclinic tangle, where 
two hyperbolic fixed points of $F$ have associated stable and unstable manifolds 
which intersect in a number of points, forming a set of lobes that protrude from one region to the other.
At time $t_0$, lobes $A_1$ and $A_2$ belong to region $A$ and lobes $B_1$ and $B_2$ to region $B$. Particles trapped in
each lobe cannot cross their bordering manifolds, but as time goes by the dynamic manifolds are transported and deformed by the flow,
since they are material lines.
After one period, lobe $A_1$ is mapped onto the lobe marked as $F(A_1)$, which belongs to region $B$. 
The same happens to lobe $A_2$, which is mapped onto $F(A_2)$. Further
iterations of the Poincar\'e map $F$ may cause lobes $F(A_1)$ and $F(A_2)$ to be deeply immersed into region $B$.
Similarly, lobes $B_1$ and $B_2$ in region $B$ are mapped to lobes $F(B_1)$ and $F(B_2)$ in region $A$.

\section{The model}

A compressible isothermal gas is considered, with constant sound speed $c_s$, 
constant dynamical viscosity $\mu$, constant magnetic diffusivity 
$\eta$, and constant magnetic permeability $\mu_0$. 
The following set of compressible MHD equations is solved 

\begin{eqnarray}
& &\partial_t \ln\rho+\mathbf{u}\cdot\nabla\ln\rho+\nabla\cdot\mathbf{u}=0,\label{eq continuity}\\
& &\partial_t\mathbf{u}+\mathbf{u\cdot}\nabla\mathbf{u}=-\nabla p / \rho + \mathbf{J\times B}/\rho+ 
    (\mu/\rho)\left(\nabla^{2}\mathbf{u}+\nabla\nabla\cdot\mathbf{u}/3\right)+\mathbf{f},\label{eq momentum}\\
& &\partial_t\mathbf{A}=\mathbf{u\times B}-\eta\mu_{0}\mathbf{J}\label{eq induction},
\end{eqnarray}
where $\rho$ is the density, $\mathbf{u}$ is the fluid velocity, $\mathbf{A}$ is the magnetic vector potential,
 $\mathbf{J} = \nabla \times \mathbf{B}/\mu_0 $ is the current density, $p$ is the 
pressure,  
$\mathbf{f}$ is an external forcing, and $\nabla p / \rho = c_s^2\nabla\ln\rho$, 
where  $c_s^2 = \gamma p/\rho$ is assumed to be constant. 
Nondimensional units are adopted, 
such that $c_s = \rho_0 = \mu_0 = 1$, where $\rho_0=\left<\rho\right>$ is the spatial average of $\rho$. 
Equations (\ref{eq continuity})--(\ref{eq induction}) are solved with the 
{\small PENCIL CODE} \footnote{http://pencil-code.googlecode.com}
in a box with 
sides $L = 2\pi$ and periodic boundary conditions, so the smallest wavenumber is $k_1=1$. The time unit is $(c_sk_1)^{-1}$
and the unit of viscosity $\nu$ and magnetic diffusivity $\eta$ is $c_s/k_1$. 
The initial conditions are $\ln \rho = \mathbf{u} = 0$, 
and $A$ is a set of normally distributed, uncorrelated random numbers with zero mean
and standard deviation 
equal to $10^{-3}$. The forcing function $\mathbf{f}$ is given by the strongly helical ABC flow,

\begin{equation}
\mathbf{f(x)}=\frac{A_{f}}{\sqrt{3}}
[\sin k_{f}z+\cos k_{f}y, \sin k_{f}x+\cos k_{f}z, \sin k_{f}y+\cos k_{f}x],
\end{equation}
where $A_f$ is the amplitude and $k_f$ the wavenumber of the forcing function.

Following Rempel et al. \cite{rempel09,rempel11}, we use $A_f = 0.1$, $k_f=5$, and
the numerical resolution varies between $64^3$ and $128^3$.  
Spatial averages are denoted by $\left< \cdot \right>$ and 
time averages by $\left< \cdot \right>_t$. References to kinetic $(Re)$ and magnetic $(Rm)$ Reynolds numbers
are based on the forcing scale 

\begin{equation}
Re = \lambda_fU/\nu, \qquad Rm = \lambda_fU/\eta, 
\end{equation}
where $\nu = \mu / \rho_0$ is the average kinematic viscosity,
$\lambda_f=2\pi/k_f$ is the forcing spatial scale, and $U=\left<u^2\right>^{1/2}$ is the mean velocity
at a time when the magnetic field is saturated. The turnover time $\tau=\lambda_f/u_{rms}$ varies between 
$\tau \approx 3$ and $\tau \approx 4.5$ for our range of $\eta$.

\section{Results}

\subsection{Bifurcation diagrams}

\begin{figure}[ht]
\begin{center}
 \includegraphics[width=0.7\columnwidth]{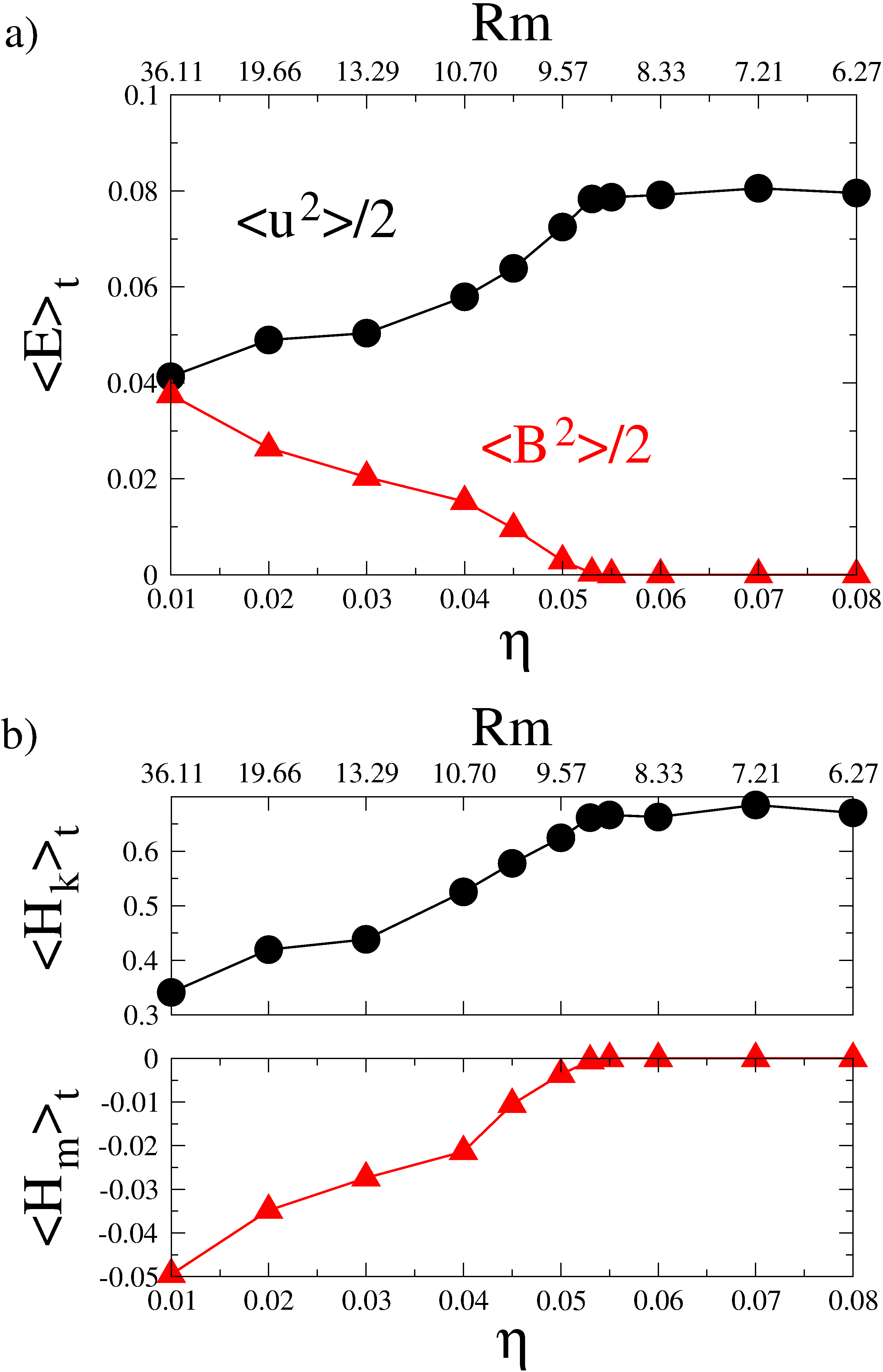}
\end{center}
 \caption{\label{fig bifs} (Color online) Bifurcation diagrams as a function of $\eta$: 
(a) kinetic (black circles) and magnetic (red triangles) energies;
(b) kinetic (black circles) and magnetic (red triangles) helicities.}
 \end{figure}

We choose $\eta$ as the control parameter and fix $\nu=0.005$, which in the absence of magnetic fields
corresponds to a spatiotemporally chaotic flow with $Re \approx 100$.
Figure \ref{fig bifs}(a) shows the bifurcation diagrams for the time--averaged magnetic ($\left<E_m\right>_t$, red triangles) 
and kinetic ($\left<E_k\right>_t$, black circles) energies as a function of $\eta$ (lower axes) or $Rm$ (upper axes). 
Averages are computed after an initial transient is dropped.
For large values of $\eta$, the seed magnetic field 
decays rapidly and there is no dynamo. 
At the  onset of dynamo action at  $\eta \sim 0.053$ ($Rm \sim 9.5$), 
the magnetic energy starts to grow at the 
expense of kinetic energy, until it saturates. 
Figure \ref{fig bifs}(b) shows in the upper panel the time--averaged kinetic helicity, $H_k = \left< \mathbf{u}\cdot\mathbf{\omega} \right>$,
where $\mathbf{\omega}=\nabla \times \mathbf{u}$ is the vorticity,
and in the lower panel the time--averaged magnetic helicity, $H_m = \left< \mathbf{A} \cdot \mathbf{B} \right>$.
For helically forced flows, the magnetic helicity is expected to have
the same sign as the kinetic helicity in scales smaller than the energy injection scale and 
the opposite sign in larger scales \cite{alexakis}. Most of the magnetic helicity in our simulations is concentrated in large scales,
as happens with the magnetic energy \cite{rempel09}, thus, $H_m$ has the opposite sign as $H_k$ in Fig. \ref{fig bifs}(b).
These quantities are crucial for the emergence of a large--scale mean field, as they are related to the 
$\alpha$--effect in mean--field dynamo theory, which is responsible for the generation 
of a mean electromotive force along the mean magnetic field by turbulent fluctuations of the velocity and magnetic fields
\cite{moffatt78,krauseradler80}. The presence of kinetic helicity is thought to be responsible
for the inverse transfer of magnetic energy from small scales to large scales, as well as the inverse transfer
of magnetic helicity from the energy injection scale to larger scales \cite{alexakis}.
In Fig. \ref{fig bifs}(b), $H_k$ is high
for large values of $\eta$  and $H_m$ is null, since there is no dynamo and a maximally helical (Beltrami) forcing
is applied to the flow. After the onset of dynamo, the magnetic field starts to contribute to the flow dynamics through the Lorentz force 
(second term at the right in Eq. \ref{eq momentum}) and $H_k$
decreases with $\eta$, as $\left|H_m\right|$ grows.

\begin{figure}[ht]
\begin{center}
 \includegraphics[width=0.7\columnwidth]{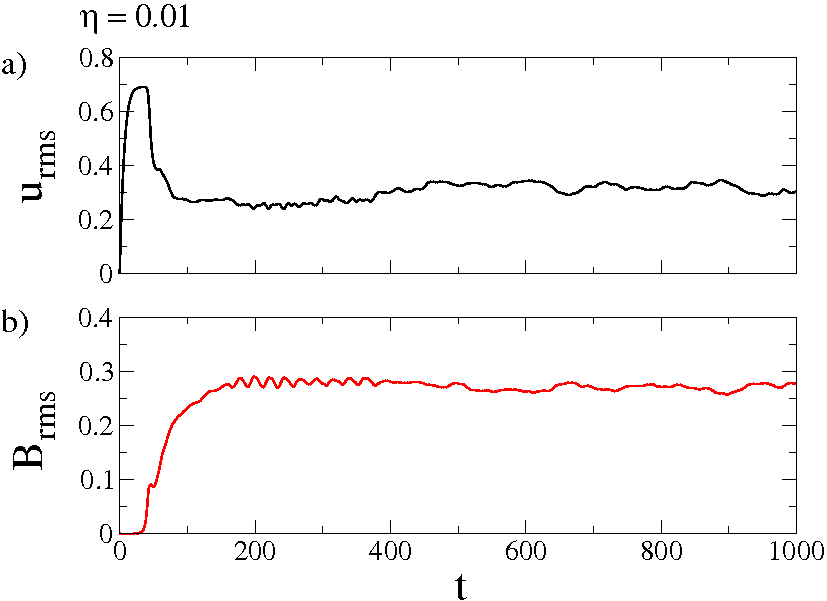}
\end{center}
 \caption{\label{fig ts01} (Color online) Time series for $u_{\rm rms}$ (upper panel) and $B_{\rm rms}$ (lower panel) for $\nu=0.005$ and $\eta=0.01$.}
 \end{figure}

\begin{figure}[ht]
\begin{center}
 \includegraphics[width=0.7\columnwidth]{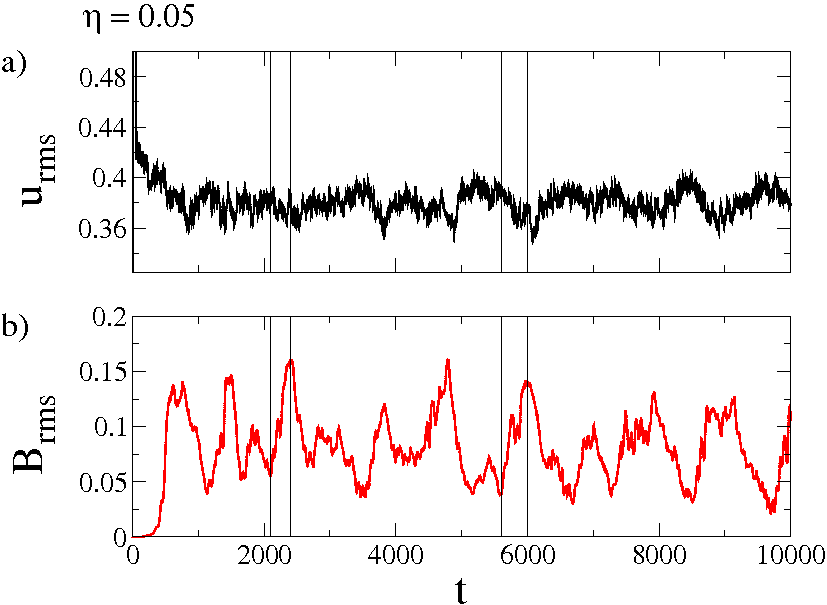}
\end{center}
 \caption{\label{fig ts05} (Color online) Time series for $u_{\rm rms}$ (upper panel) and $B_{\rm rms}$ (lower panel) for $\nu=0.005$ and $\eta=0.05$.}
 \end{figure}

We focus on two values of $\eta$. For $\eta=0.01$ the magnetic field is close to equipartition after saturation, 
as seen in the comparison between the time series of $B_{\rm rms}$ and $u_{\rm rms}$ in Fig. \ref{fig ts01}.
For $\eta=0.05$, close to the onset of dynamo action, the magnetic energy is almost an order of magnitude smaller than the kinetic energy and
 the time series are strongly intermittent, as shown in Fig. \ref{fig ts05}. Two pairs of vertical lines
in Fig. \ref{fig ts05} mark the beginning and apex of two bursts of magnetic energy around times $t=2400$ and $t=6000$.
This is a type of on--off intermittency due to a blow--out bifurcation, as discussed by Rempel et al. \cite{rempel09}.

\begin{figure*}[ht]
\begin{center}
 \includegraphics[width=\columnwidth]{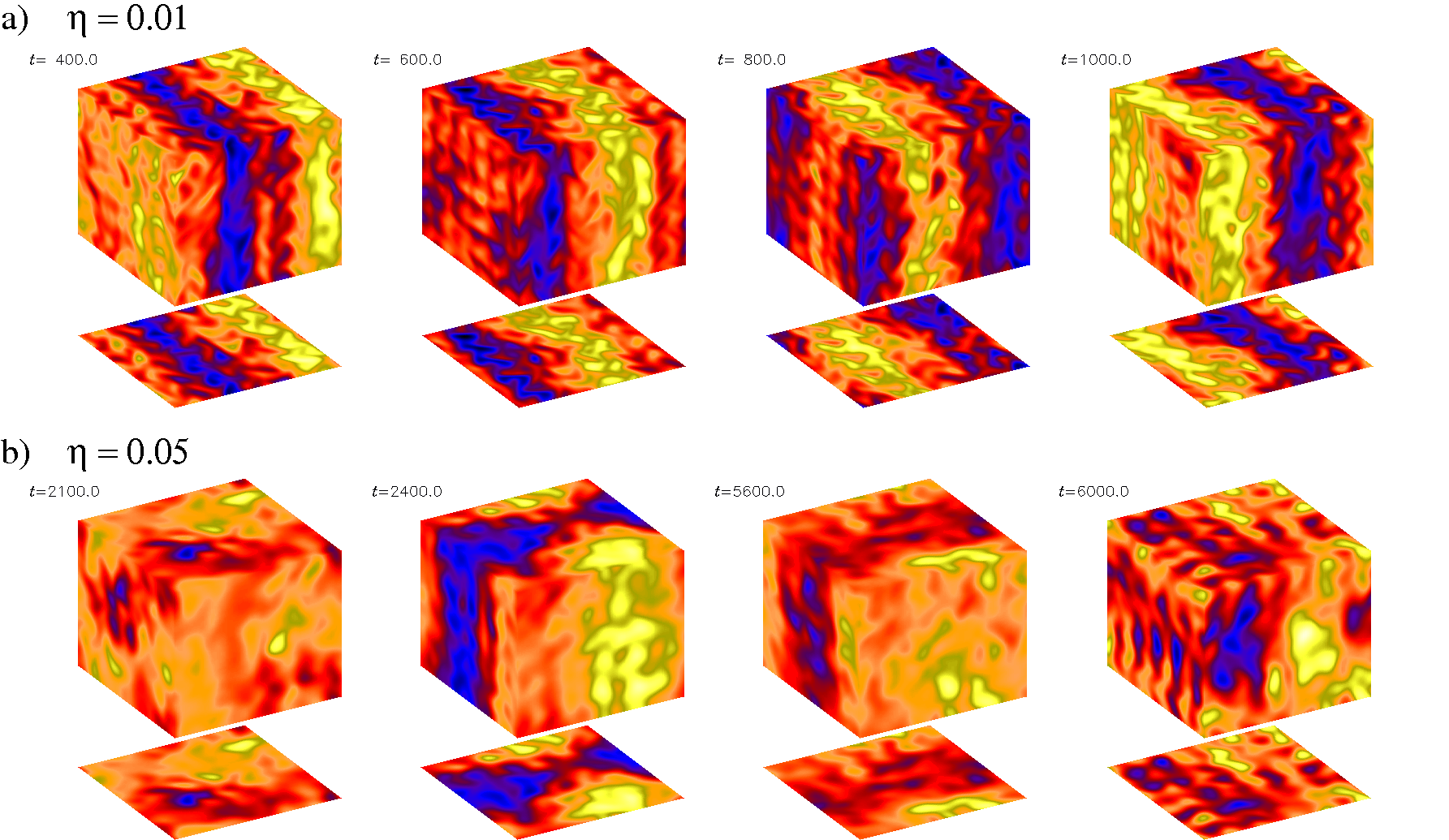}
\end{center}
 \caption{\label{fig patts} (Color online) (a) Intensity plots of $B_z$ at four different times, 
showing the evolution of a large-scale coherent pattern modulated along the $x$ direction for $\eta=0.01$; 
(b) same as (a) but for $\eta=0.05$, showing intermittent switching between ordered and disordered patterns.}
 \end{figure*}

The magnetic field structures are depicted in Fig. \ref{fig patts} for the two values of $\eta$ and different times. For $\eta=0.01$
(upper panel) there is a robust coherent large--scale $B_z$ component accompanied by small--scale
turbulent fluctuations. For $\eta=0.05$ (lower panel), the magnetic field displays an intermittent switching
between coherent and incoherent large-scale structures [Fig. \ref{fig patts}(b)] and there is no preferred direction
for field alignment.

\subsection{Lagrangian analysis}

\begin{figure*}[h]
\begin{center}
 \includegraphics[width=\columnwidth]{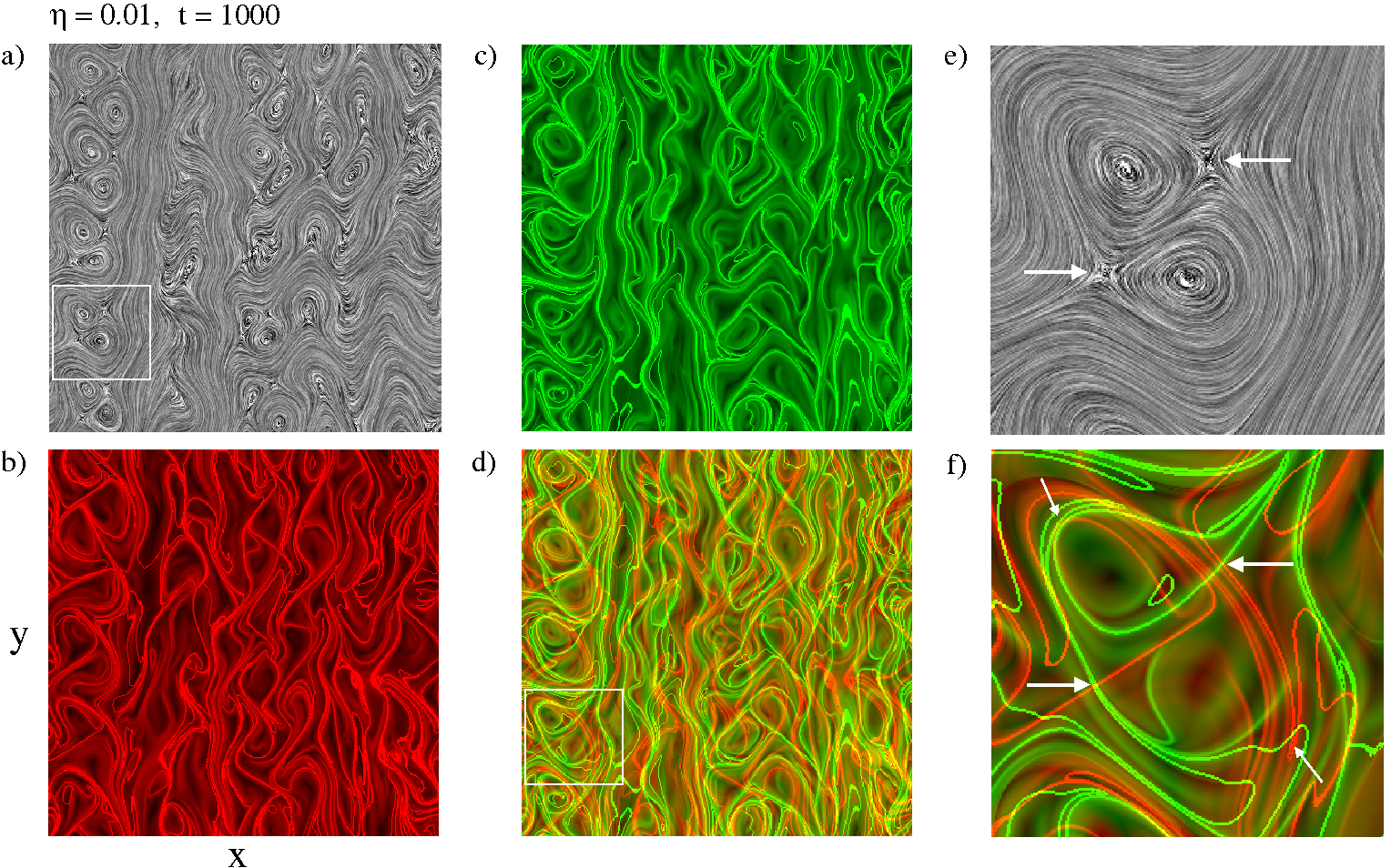}
\end{center}
 \caption{\label{fig lcs2} (Color online) (a) Line integral convolution plot showing the streamlines
of the $xy$--components of the velocity field at $t=1000$ for $\eta=0.01$; (b) the attracting Lagrangian
coherent structures (LCS) (red); (c) the repelling LCS (green); (d) superposition of (b) and (c); (e) enlargement of the 
square region of (a); (f) enlargement of the square region of (d). }
 \end{figure*}

The contrast between the Eulerian and Lagrangian analyses is depicted in Figure \ref{fig lcs2} for $\eta =0.01$,
when the magnetic field has settled to a spatially regular mean field. Figure \ref{fig lcs2}(a) shows
the line integral convolution (LIC) plot \cite{cabral93} for the velocity field at $z=0$. The LIC plot
reveals the streamlines of the $(u_x, u_y)$ velocity components on this plane at $t=1000$. Arrays of counter--rotating vortices
found in the ABC--flow can still be seen, intermixed with long streaks and displaced vortices. 
To obtain the Lagrangian coherent structures we compute the maximum FTLE field. 
For the attracting LCS (finite--time unstable manifolds) we need to integrate
the compressible MHD equations backward in time, which is a major problem, since the system is dissipative. 
We resort to interpolation of recorded data to compute these fields. A run of Eqs. (\ref{eq continuity})--(\ref{eq induction}) 
from $t_0-\tau$ to $t_0+\tau$
is conducted and full three-dimensional snapshots of the velocity fields are saved at each $dt=0.5$ time interval.
Following \cite{padberg07} and \cite{swinney}, linear interpolation in time and third order Hermite interpolation in space is used to obtain
the continuous set of vector fields necessary to obtain the particle trajectories by integration of Eq. (\ref{eq odes}). 
For backward integration, $\mathbf{\dot{x}}=-\mathbf{u}(\mathbf{x},t)$
is solved instead, as snapshots are read from $t_0$ to $t_0-\tau$. 
Figure \ref{fig lcs2}(b) shows
the backward--time maximum FTLE field computed with $\tau=-10$ and $t_{0}=1000$. Bright colors correspond 
to large values of $\sigma_1$ and dark regions to low values. The ridges seen as bright 
red lines approximate the attracting LCS. Figure \ref{fig lcs2}(c) shows the 
forward--time maximum FTLE field for $\tau=10$, whose ridges provide the repelling LCS. Figure \ref{fig lcs2}(d) is a superposition of (b) and (c),
and represents the so--called ``Lagrangian skeleton of turbulence" \cite{swinney}.
Figures \ref{fig lcs2}(e) and \ref{fig lcs2}(f) are enlargements of the square regions in Figs. \ref{fig lcs2}(a) and \ref{fig lcs2}(d), 
respectively. Note that the LIC plot of the velocity field 
in Fig. \ref{fig lcs2}(e) shows a structure similar to the homoclinic connections of Fig. \ref{fig tangles}(b).
Here, the arrows point to two hyperbolic stagnation points in the $(u_x, u_y)$ field. On the other hand, when one moves to the Lagrangian frame
(Fig. \ref{fig lcs2}(f)) the picture becomes much more complex, with a number of homoclinic and heteroclinic crossings, 
as in Fig. \ref{fig tangles}(c). The two larger arrows point to the same location of the 
stagnation points. The smaller arrows indicate two lobes that cross other LCS and permit the transport of particles between
vortices through lobe dynamics.
The LCS in Fig. \ref{fig lcs2} were computed using
$384 \times 384$ fiducial particles uniformly distributed on the plane $z=0$. 
For each fiducial particle, the trajectories of six near neighboring particles are computed to obtain the 
deformation gradient by second-order centered finite--differences.

\begin{figure}[h]
\begin{center}
 \includegraphics[width=0.5\columnwidth]{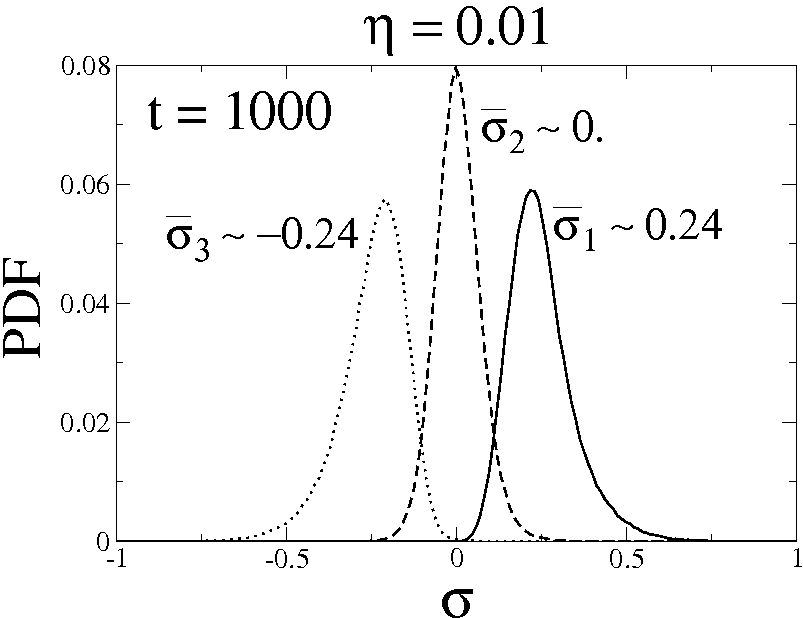}
\end{center}
 \caption{\label{fig hist01} PDFs of the finite--time Lyapunov exponents at $t_0=1000$ and $\tau=10$ for $\eta=0.01$.}
 \end{figure}

In order to quantify the degree of particle dispersion or chaotic mixing in the flow,
Fig. \ref{fig hist01} shows the probability distribution functions (PDFs) of the three finite--time Lyapunov exponents for the same
state shown in Fig. \ref{fig lcs2}. The PDFs were obtained from a set o $64^3$ initial conditions
uniformly distributed in the box at $t_0=1000$, with $\tau=10$. There is a considerable amount of 
trajectories with two positive Lyapunov exponents,
thus sheet--like magnetic field structures are expected. The broad tails in $\sigma_1$ are due to initial conditions
that are very close to the repelling material lines, which are regions where the stretching is stronger than the average.
On the other hand, the broad tails in negative values of $\sigma_3$ reflect contraction in the vicinity of 
the attracting material lines. Since the flow is weakly compressible, with the Mach number below 0.4,
for almost all initial conditions one of the exponents is close to zero. 
The PDF for $\sigma_2$ shows a Gaussian distribution. The overbars on $\sigma$ denote average values.

\begin{figure}[h]
\begin{center}
 \includegraphics[width=1.\columnwidth]{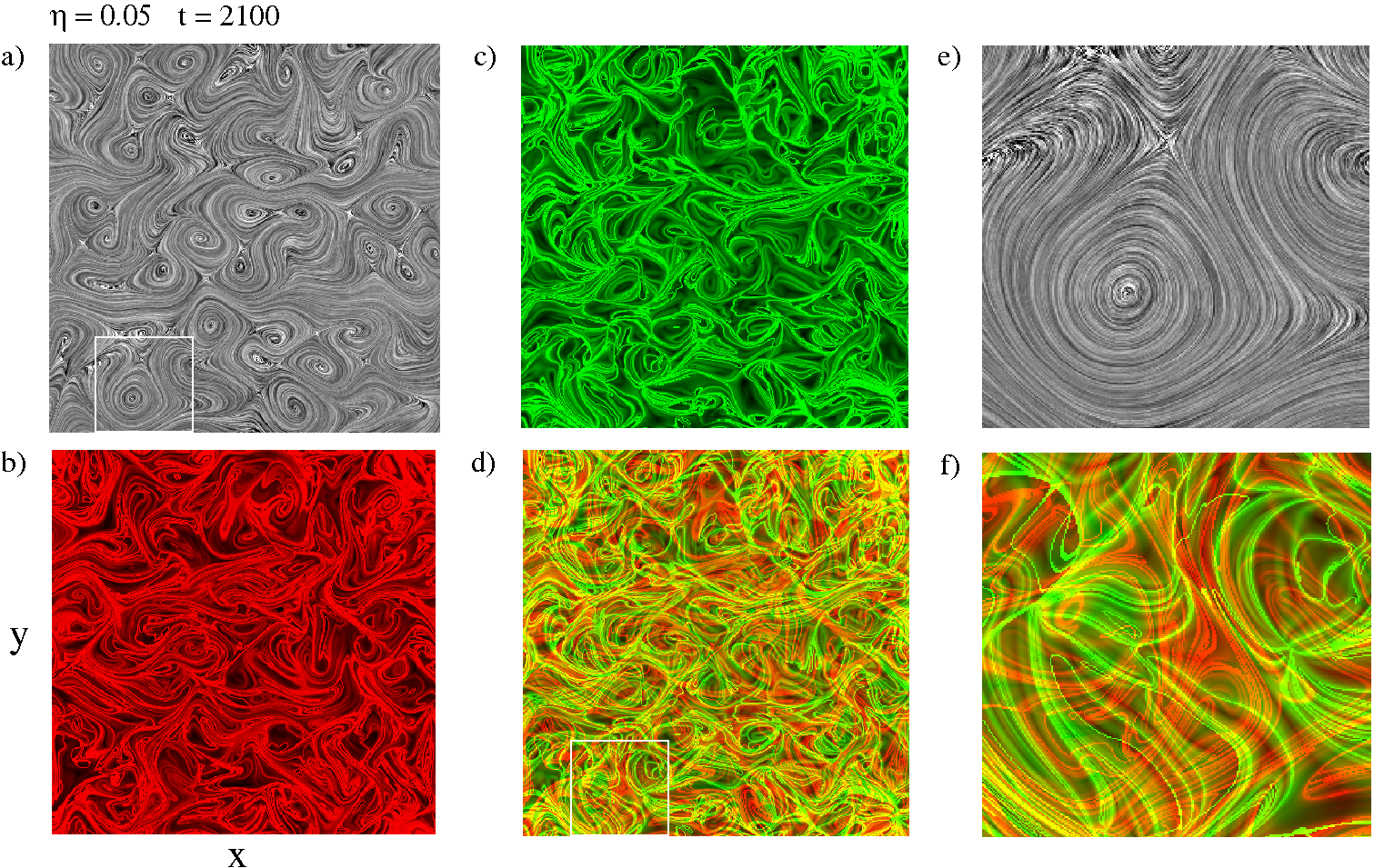}
\end{center}
 \caption{\label{fig lcs2100} (a) Line integral convolution plot showing the streamlines
of the $xy$--components of the velocity field at $t=2100$ for $\eta=0.05$; (b) the attracting Lagrangian
coherent structures (LCS) (red); (c) the repelling LCS (green); (d) superposition of (b) and (c); (e) enlargement of the 
square region of (a); (f) enlargement of the square region of (d).}
 \end{figure}

For $\eta =0.05$ the time series of $B_{\rm rms}$ and $u_{\rm rms}$ are intermittent. To understand the influence of
{\bf B} on {\bf u}, the FTLE are computed for different initial times marked with vertical lines in Fig. \ref{fig ts05}.
Figure \ref{fig lcs2100} shows the LIC and LCS plots at $t=2100$, just before a burst of magnetic energy in the
time series of Fig. \ref{fig ts05}(b). 
In comparison to $\eta=0.01$, there seems to be less order in the distribution of vortices in the LIC plot of Fig. \ref{fig lcs2100}(a)
than in \ref{fig lcs2}(a) and the greater complexity in the distribution
of material lines in the LCS plots of Figs. \ref{fig lcs2100}(b)--(d) and (f) indicates that transport of passive scalars is enhanced 
due to the frequent crossings of attracting and repelling lines. This is expected, since $B_{\rm rms}$ is much lower 
for $\eta=0.05$ and has a smaller impact on the velocity field.

\begin{figure}[h]
\begin{center}
 \includegraphics[width=1.\columnwidth]{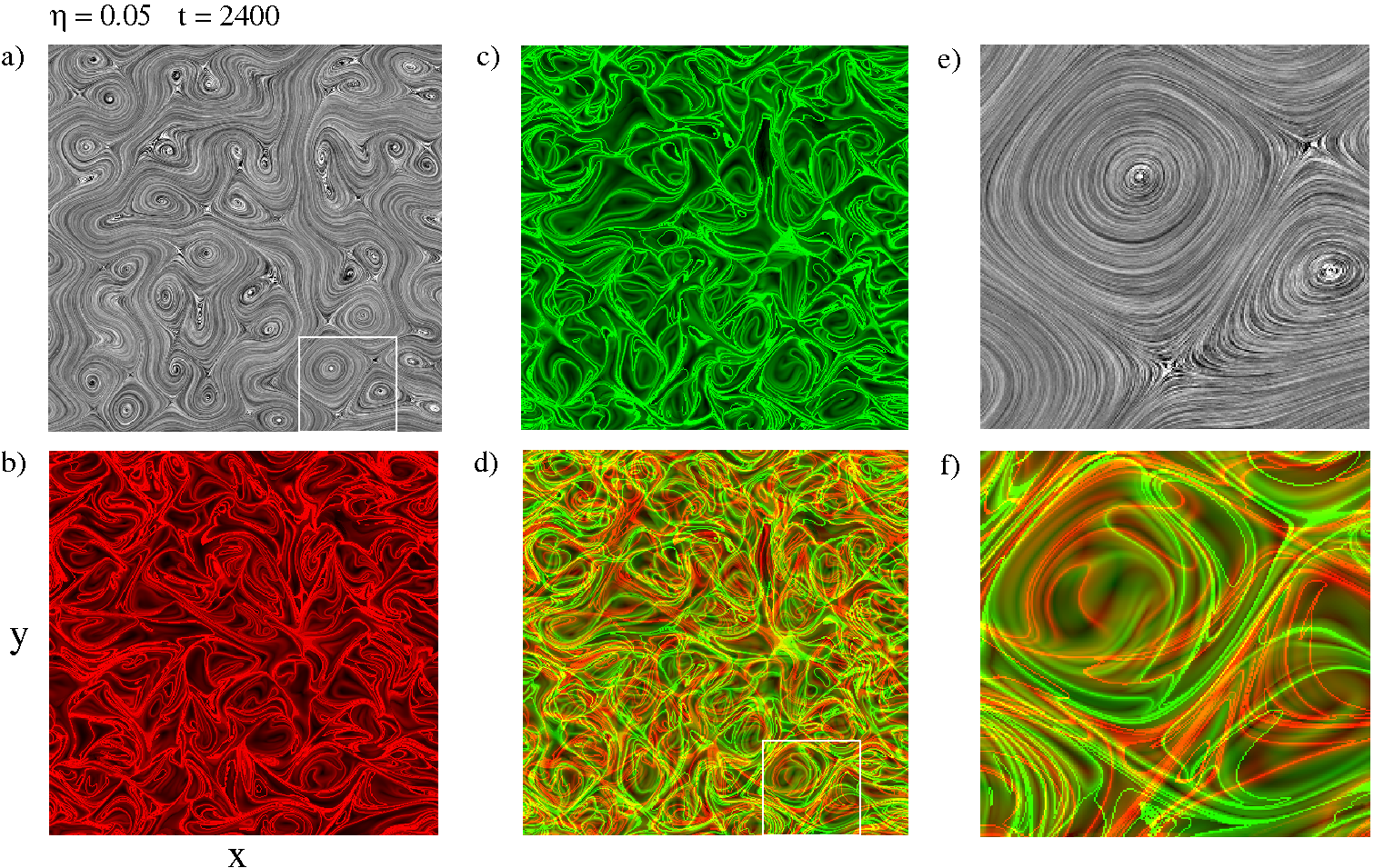}
\end{center}
 \caption{\label{fig lcs2400} (a) Line integral convolution plot showing the streamlines
of the $xy$--components of the velocity field at $t=2400$ for $\eta=0.05$; (b) the attracting Lagrangian
coherent structures (LCS) (red); (c) the repelling LCS (green); (d) superposition of (b) and (c); (e) enlargement of the 
square region of (a); (f) enlargement of the square region of (d).}
 \end{figure}

\begin{figure}[h]
\begin{center}
 \includegraphics[width=0.5\columnwidth]{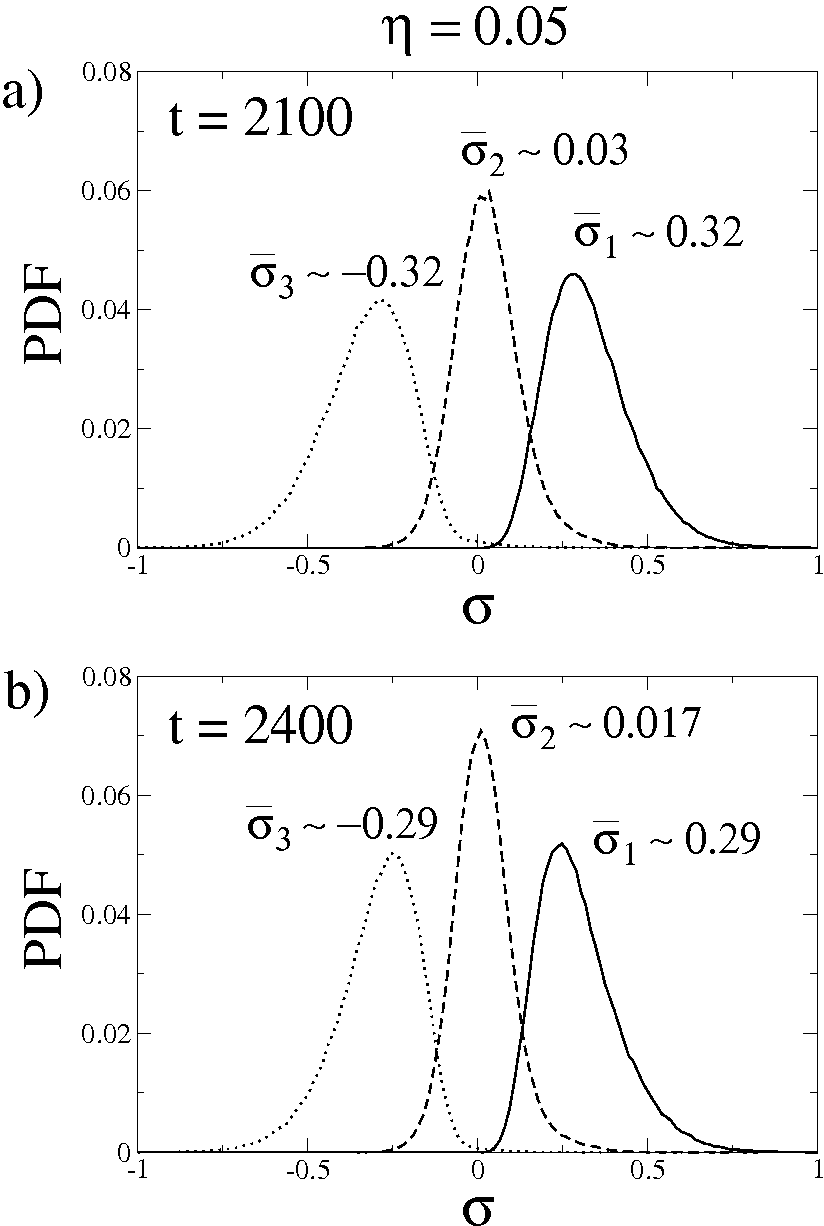}
\end{center}
 \caption{\label{fig hist05} PDFs of the finite--time Lyapunov exponents for $\eta=0.05$ at (a) $t=2100$ and (b) $t=2400$.}
 \end{figure}

At $t=2400$ the time series of $B_{\rm rms}$ has a peak
of energy burst. As seen in Fig. \ref{fig lcs2400}, there is a stronger impact of this magnetic energy release 
on the velocity field in comparison with t=2100 (Fig. \ref{fig lcs2100}).
The LIC plot of Fig. \ref{fig lcs2400}(a) does not show much difference in relation to Fig. \ref{fig lcs2100}(a).
However, the LCS plots of Figs. \ref{fig lcs2400}(b)--(d) show wider regions of low particle dispersion. 
This is clearer in Fig. \ref{fig lcs2400}(f), which shows an intermediate level of complexity in 
comparison to Figs. \ref{fig lcs2}(f) and \ref{fig lcs2100}(f).
Thus, a stronger magnetic
field diminishes chaotic mixing in the velocity field, which is measured by the PDFs of the FTLE, 
shown in Fig. \ref{fig hist05}.

\begin{figure}[h]
\begin{center}
 \includegraphics[width=0.5\columnwidth]{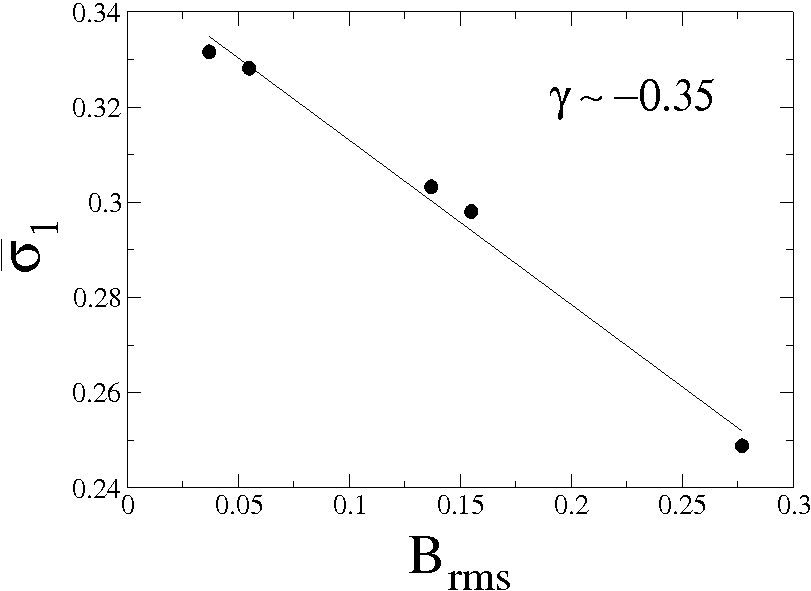}
\end{center}
 \caption{\label{fig brmsigma} Maximum finite--time Lyapunov exponent of the velocity field $(\bar{\sigma}_1)$ 
as a function of $B_{\rm rms}$ using data from Table \ref{tabela}.
The fitted line has slope $\gamma \sim -0.345$.}
 \end{figure}

One can see that the PDFs for the intermittent dynamo (Fig. \ref{fig hist05}(a),(b)) are wider than for the regular mean--field
dynamo (Fig. \ref{fig hist01}). They also have a larger $\bar{\sigma}_1$,
revealing that the flow is more chaotic for $\eta=0.05$ than for $\eta=0.01$. Moreover,
broader tails in the PDFs of Fig. \ref{fig hist05} mean that more intermittency is to be expected in the 
evolution of passive scalars at $\eta=0.05$. 
A summary of the results is found in Table \ref{tabela},  
which shows $\bar{\sigma}_{1,2,3}$ and their standard deviations for $\eta=0.01$ at $t=1000$ and 
for $\eta=0.05$ at four values of $t$ representing the beginning and apex of the 
two magnetic energy bursts indicated in Fig. \ref{fig ts05}. 
The mean value $\bar{\sigma}_1$ and the standard deviation decrease at both bursts.
Figure \ref{fig brmsigma} is a plot of $\bar{\sigma}_1$ as a function of $B_{\rm rms}$ using only data from Table \ref{tabela}, 
which are fitted with the linear equation $\bar{\sigma}_1 \approx 0.348 - 0.345 B_{\rm rms}$.
Although we need more statistics to draw conclusions, our preliminary results suggest that
the decay of chaoticity in the velocity field is proportional to $B_{\rm rms}$.

\begin{table}
\caption{\label{tabela}Mean values and standard deviations of finite--time Lyapunov exponents.}
\begin{tabular*}{\textwidth}{@{\extracolsep{\fill}}  l | c | c  c c c }
\br
 & $\eta=0.01$ &  $\eta=0.05$ & & &\\
\hline
 & $t=1000$ & $t=2100$ & $t=2400$ & $t=5600$ & $t=6000$ \\
\hline
$\bar{\sigma}_1$ & 0.249 & 0.328 & 0.298 & 0.332 & 0.303\\
$\bar{\sigma}_2$ & 0.006 & 0.030 & 0.017 & 0.032 & 0.021\\
$\bar{\sigma}_3$ & -0.245 & -0.328 & -0.292 & -0.333 & -0.299\\
std$(\sigma_1) \quad$ & 0.096 & 0.125 & 0.123 & 0.125 & 0.124\\
std$(\sigma_2) \quad$ & 0.067 & 0.098 & 0.089 & 0.098 & 0.091\\
std$(\sigma_3) \quad$ & 0.098 & 0.137 & 0.124 & 0.138 & 0.127\\
\br
\end{tabular*}
\end{table}

\section{Conclusions}

We have employed Lagrangian coherent structures (LCS) and the statistics of finite--time Lyapunov exponents (FTLE)
to study chaotic stirring in 3--D MHD dynamo simulations with helical forcing. Attracting LCS provide the pathways that
are more likely to be followed by passive scalars and their crossings with repelling LCS provide the 
mechanism for transport between different regions of the fluid. The PDFs of FTLE provide a quantification
of chaotic mixing in the flow. We explored the impact of the magnetic field on the velocity field in a saturated nonlinear dynamo and in an 
intermittent dynamo, and the maximum FTLE was shown to be a linear function of the magnetic energy.
The increase in the flow's chaoticity when the magnetic diffusivity is increased from $\eta=0.01$ to $\eta=0.05$
is the result of the reduction of the effect of the Lorentz force upon the velocity
field. Enhanced chaoticity leads to stronger line stretching and field amplification, and the 
``competition" between this effect and destruction of magnetic flux due to magnetic diffusion seems to be the main cause of the intermittent 
time series of magnetic energy observed when $\eta=0.05$, which is close to the critical value for dynamo action.

Our analysis has direct applications in astrophysics, where the equipartition--strength magnetic fields observed in planets and stars 
are thought to be the result of a dynamo process, whereby kinetic energy from the motion 
of a conducting fluid is converted into magnetic energy \cite{axelPR}.
Experimental detection of LCS and computation of FTLE in the solar surface can be performed using velocity fields estimated from
observational data. Such estimations can be obtained 
from digital images using the optical flow algorithm, employed by \cite{colaninno06} to
extract the velocity field from images of coronal mass ejections obtained with the SOHO LASCO C2 coronagraph. 
Recently, horizontal velocity
fields in the photosphere were inferred from Hinode images \cite{tan09} and the Swedish Vacuum Solar Telescope \cite{getling10}.
Solar subsurface flows can be inferred from helioseismic data \cite{woodard02}, thus LCS
can also aid the tracing of particle transport by turbulence in stellar interiors.

\ack{
 A.C.L.C. and E.L.R. acknowledge support from CNPq (Brazil) and CAPES (Brazil). E.L.R. acknowledges support from FAPESP (Brazil) and NORDITA (Sweden).
A.C.L.C. acknowledges the award of a Marie Curie
International Incoming Fellowship and the hospitality of Paris
Observatory. 
}

\end{document}